\documentclass[a4paper,12pt,epsf]{article}
\usepackage{epsfig}
\usepackage{amsmath}
\usepackage{amscd}
\usepackage{amsthm}
\usepackage{amssymb}
\def\fbi{~\mbox{fb$^{-1}$}}
\def\fb{~\mbox{fb}}

\def\rhot{\tilde\rho}

\def\nn{{\nonumber}}

\newcommand{\be}{\begin{equation}}
\newcommand{\ee}{\end{equation}}
\newcommand{\ba}{\begin{eqnarray}}
\newcommand{\bea}{\begin{eqnarray}}
\newcommand{\eea}{\end{eqnarray}}
\newcommand{\ea}{\end{eqnarray}}

\newcommand{\nc}{\newcommand}
\nc{\postscript}[2]
{\setlength{\epsfxsize}{#2\hsize}\centerline{\epsfbox{#1}}}
\nc{\non}{\nonumber}
\nc{\hc}{\hbox {h.c.}} \nc{\re}{\hbox {Re}} 
\nc{\mev}{\hbox {MeV}} \nc{\gev}{\;\hbox {GeV}} \nc{\tev}{\;\hbox {TeV}}
\def\lsim{\mathrel{\raise.3ex\hbox{$<$\kern-.75em\lower1ex\hbox{$\sim$}}}}
\def\gsim{\mathrel{\raise.3ex\hbox{$>$\kern-.75em\lower1ex\hbox{$\sim$}}}}

\begin{document}
\baselineskip=20pt

\vspace*{-2cm}
\begin{flushright}
$\vcenter{
\hbox{DFF-405-5-03}
}$
\end{flushright}

\begin{center}
  {\Large  \bf The scalar sector in an extended electroweak
gauge symmetry model}

\vspace{.5cm} { Stefania DE CURTIS$^a$,
Donatello DOLCE$^{b}$, Daniele DOMINICI$^{a,b}$ }

\emph{ $^a$ INFN, Sezione di Firenze, Italy.}

\emph{ $^b$ Dip. di Fisica, Univ. degli Studi di Firenze, Italy.}
\end{center}

\vspace{-.5cm}
\noindent
\rule{\textwidth}{.1mm}
\begin{abstract}
\noindent The scalar sector of the linear formulation of the
degenerate BESS model is analyzed. The model predicts two
additional scalar states which mix with the SM Higgs. As a
consequence  the properties of the SM Higgs are modified and Higgs
precision measurements can constrain the mixing angle. One of the
two additional Higgses has no coupling to fermions and suppressed
couplings to ordinary gauge bosons, therefore its detection is
difficult.
  The production  of the other two
 Higgses at future $e^+e^-$ linear colliders
 in the
Higgstrahlung and  fusion channels  is investigated.
\end{abstract}
\vspace{-.5cm}
\rule{\textwidth}{.1mm}
{\footnotesize
\vspace{-.2cm}
}

\section{Introduction}
There has been recently a renewed interest in models with extended
electroweak symmetry in the context of little Higgs models
(for a review see \cite{Schmaltz:2002wx}). Their
 low energy description is based on effective lagrangians
constructed using extended gauge symmetries including, in general,
copies of $SU(2)$ and $U(1)$ groups. Similar gauge symmetry
structures  also appear in   effective lagrangians for technicolor
and  non commuting extended technicolor \cite{Chivukula:1996gu}.
The degenerate BESS model \cite{Casalbuoni:1995yb} is a non linear
description based on the gauge symmetry group ${\cal G} = SU(2)_L
\otimes U(1)_Y \otimes SU(2)^\prime_L \otimes SU(2)^\prime_R $
 and therefore can be used 
as a general parameterization of classes of models. New
vector gauge bosons are introduced and, in order to include  the possible
(composite) scalar fields, a linear formulation was  proposed
\cite{Casalbuoni:1996wa}, describing  the
 breaking of the group $\cal G$ at some
high energy scale $u$ to $SU(2)_{weak} \otimes U(1)$ and finally, at the electroweak scale $v$, to $U(1)_{em}$.
The model, in the limit of large $u$,
 gives back the Standard Model (SM) with a light Higgs
and the contributions to the $\epsilon$ (or $S,T,U$) parameters are $O((v/u)^2 s_\varphi^2)$,
being $\varphi$ the mixing angle of the charged gauge boson sector.
As a consequence
 the model is only weakly constrained by the electroweak precision
measurements. The phenomenology of the additional gauge vector
bosons has been already addressed: detection of new vector resonances
will be possible at the LHC up to masses of approximately 2 TeV
for $s_\varphi \sim 0.14$ \cite{Casalbuoni:2000gn}.
Aim of this paper
is to perform the analysis of the scalar sector of the model and
study its properties at future linear colliders (LC's) which offer the possibility
of detecting the Higgs and
also  performing precision measurements of Higgs boson cross sections,
partial widths and of the trilinear Higgs coupling. The investigation of the
scalar sector at the LHC will be the subject of a separate paper.

In Section 2 we review the scalar sector of the linear degenerate BESS
model
and derive the couplings of the scalars to fermions, gauge bosons and
their self-interactions. In Sections 3 and 4 analytical and numerical results
are obtained for the widths of the Higgses and the production cross sections
at future LC's. In Section 5 we study the bounds on the
parameters of the scalar sector of the model from the LC measurements.

\section{The Linear BESS model: a new parameterization}
Existing experimental data confirm with great accuracy the
SM of the electroweak interactions, therefore
only extensions which smoothly modify its predictions are still
conceivable.
 There are  examples of
strong symmetry breaking  schemes, like degenerate BESS
\cite{Casalbuoni:1995yb},   satisfying this property. The model
describes, besides the standard $W^\pm$, $Z$ and $\gamma$ vector
bosons, two new triplets of spin 1 particles, $V_ L$ and $V_R$.
The interest in this scheme was due to its decoupling property:
in the limit of infinite mass of the heavy vector bosons  one
gets back the Higgsless SM.  The original philosophy of the
non-linear version was based on the idea that the non-linear
realization would be the low-energy description of some
underlying dynamics giving rise to the breaking of the
electroweak symmetry. A linear realization of this model (L-BESS)
was proposed in \cite{Casalbuoni:1996wa}. This scenario is a
possible effective description of  technicolor
 and of its  generalizations as non-commuting technicolor
models \cite{Chivukula:1996gu}, where  an underlying
strong dynamics produces heavy Higgs composite particles. The
L-BESS model
 describes the theory, as a renormalizable theory,
 at the level of its composite
states, vectors (the new heavy bosons), and scalars (Higgs
bosons).

Let us first review the main properties of
the  L-BESS model, in particular of its scalar sector, on which we
will focus in the present study.
The model is a $SU(2)_L\otimes U(1)_Y \otimes SU(2)_{L}^\prime
\otimes SU(2)_{R}^\prime$ gauge theory, breaking at  some high
scale $u$ to $SU(2)_{weak}\otimes U(1)$ and  breaking again at the electroweak scale $v$ to $U(1)_{em}$.

The L-BESS model contains,  besides the standard Higgs sector described by
the field $\tilde U$, two additional scalar fields $\tilde L$ and
$\tilde R$. They  belong to the (2,2,0,0), (2,0,2,0) and (0,2,0,2)
 representations of the global symmetry group $G=SU(2)_L\otimes
SU(2)_R\otimes SU(2)_{L}^\prime \otimes SU(2)_{R}^\prime$,
respectively. The two breakings are induced by the vacuum expectation
values $\langle\tilde L\rangle= \langle\tilde R\rangle=u$ and
$\langle\tilde U\rangle=v$. We will assume $u\gg v$. Proceeding
in the  standard way, we  build up the kinetic terms for the fields
in terms of the covariant derivatives with respect to the local
$SU(2)_L\otimes U(1)_Y\otimes SU(2)_{L}^\prime \otimes
SU(2)_{R}^\prime$, by introducing  ${\vec V}_L~({\vec V}_R)$ as
gauge fields of $SU(2)_L^\prime ~ (SU(2)_R^\prime)$,
 with a common gauge coupling  $g_2$,
whereas $g_0$ and $g_1$ are the gauge couplings of the $SU(2)_L$
and $U(1)_Y$ gauge groups respectively. The scalar potential
responsible for the  breaking of  the original symmetry down to
the $U(1)_{em}$ group is constructed by requiring invariance
with respect to the group
 $G$ and  the discrete symmetry ${\tilde L}\leftrightarrow {\tilde
 R}$.
 As far as the fermions are concerned they transform
as in the SM with respect to the group $SU(2)_L\otimes U(1)_Y$,
correspondingly the Yukawa terms are built up exactly as in the
SM \cite{Casalbuoni:1995yb}.

 We parameterize the scalar
fields as $\tilde L=\tilde\rho_L L$, $\tilde R= \tilde\rho_R  R$,
$\tilde U=\tilde\rho_U  U$, with ${L}^\dagger {L}=I$,
${R}^\dagger{R}=I$
 and ${U}^\dagger  {U}=I$.
The scalar potential is  expressed in terms of three Higgs
fields: \be \label{potential}
 V(\rhot_U,\rhot_L,\rhot_R)= 2 \mu^2
(\rhot_L^2+\rhot_R^2)+ \lambda (\rhot_L^4+\rhot_R^4) +2 m^2
\rhot_U^2 +
 h \rhot_U^4
+ 2 f_3 \rhot_L^2\rhot_R^2 + 2 f \rhot_U^2(\rhot_L^2+\rhot_R^2).
\ee
 We assume $m^2, \mu^2 < 0$, and   $\lambda, h > 0$ for the
vacuum stability.

From the stationarity conditions, the requirements
 $\mu^2<0$ and $m^2<0$ lead to:
 \be \label{cond:parametri:1}
(f_3 + \lambda) + f x^2 > 0\,,~~~~~~~ h x^2 + 2 f > 0\, ~~~~~{\rm
with} ~~~~~ x=v/u.\ee

After shifting the fields by their v.e.v's $u,v\neq 0$, the mass
eigenvalues are: \bea\label{masses}
 m_{\rho_L}^2 &=& 8 v^2 \frac{(\lambda - f_3)}{x^2},\nn\\
m_{\rho_R}^2 &=& 8v^2( \frac{\lambda + f_3}{x^2} c_{\alpha}^2 + h
s_{\alpha}^2 + \sqrt2 \frac{f}{x} s_{2\alpha}), \nn\\ m_{\rho_U}^2
&=& 8v^2( \frac{\lambda + f_3}{x^2} s_{\alpha}^2 + h c_{\alpha}^2
-  \sqrt2 \frac{f}{x} s_{2\alpha}),
\eea
with
\begin{equation}
\label{alfa} \tan{(2\alpha)} =  \frac{2 \sqrt2 f x}{(\lambda +
f_3 ) - h x^2}\, 
\end{equation}
By requiring positive mass
eigenvalues,  we get $\lambda-f_3>0$ and  $\lambda+f_3> 2 f^2/h$.
As a consequence, the conditions in eq. (\ref{cond:parametri:1}) are
automatically satisfied if we choose $f\ge 0$. For simplicity we
will restrict our analysis  to non negative values of the $f$
parameter.

The Higgs boson mass eigenstates are: \be\label{eigenstates}
\rho_L = \frac{1}{\sqrt2}(\tilde{\rho_L} - \tilde{\rho_R}),~~~~
{\rho}_R = \frac{c_\alpha}{\sqrt2}(\tilde{\rho_L} +
\tilde{\rho_R}) + s_{\alpha}\tilde{\rho_U}, ~~~~{\rho}_U = \frac{-
s_\alpha}{\sqrt2}(\tilde{\rho_L} + \tilde{\rho_R}) +
c_{\alpha}\tilde{\rho_U}. \ee 
Since fermions are only coupled to
$\tilde\rho_U$, the Higgs field $\rho_L$ is  not coupled to
fermions. We will refer to $\rho_U$ and $\rho_R$ as standard-like
Higgs bosons; their couplings to fermions are obtained by rescaling the
SM Higgs ones by $c_\alpha$ and $s_\alpha$ respectively.

The results in \cite{Casalbuoni:1996wa} are recovered by
taking  $x\to 0$, for  $f$, $\lambda$, $f_3$, $h$ finite.
In this limit $\alpha \simeq  \sqrt2 f x/(\lambda + f_3 )$;
$m_{\rho_L,\rho_R}^2$ grow like $1/x^2$ while $m_{\rho_U}^2$
is finite. If in addition we   turn  off the mixing between the light and
heavy scalar sector ($f=0$),  we get back the SM Higgs sector
described by $\rho_U$.

Eqs. (\ref{masses}, \ref{eigenstates}) have a $2\pi$ periodicity.
However, by inspection, it is possible to limit the study of the
properties of the scalar sector of the L-BESS model to the region
$\alpha \in [0, \pi/2]$ where $m_{\rho_R}\ge m_{\rho_U}$ for
$f\ge 0$. For different values of $\alpha$ the results are easily
obtainable by opportunely changing the role of the standard-like
Higgs fields and the value of the mixing angle.
The parameters of the scalar potential in eq. (\ref{potential})
are six: $m, \mu, \lambda, h, f$ and $f_3$. By using the minimum
conditions we can eliminate
 $m$ and $\mu$ in favor of $u$ and $v$, or equivalently of $x$
and  $v$. Furthermore, from eqs. (\ref{masses}, \ref{alfa}), by
expressing $\lambda, h, f$ and $f_3$ in terms of $\alpha$   and
the three Higgs bosons masses,  we obtain the following trilinear
couplings among the Higgs fields: \bea &&V^{tril}(\rho_U,
\rho_L,\rho_R) =\Big[ \frac{m_{\rho_U}^2}{2 v}( c^3_{\alpha} -
\frac{ x}{ \sqrt2}  s^3_{\alpha})\Big] \rho_U^3+ \Big[
\frac{m_{\rho_R}^2}{2 v}(  s^3_{\alpha}  + \frac{ x}{ \sqrt2}
c^3_{\alpha})\Big] \rho_R^3\nn\\ &&+\Big[\frac{(2 c_{\alpha} +
\sqrt2 x s_{\alpha}) s_{2 \alpha} (2 m_{\rho_U}^2 + m_{\rho_R}^2
)}{ 8 v}\Big]\rho_R\rho_U^2+\Big[\frac{x c_{\alpha}(m_{\rho_R}^2
+ 2 m_{\rho_L}^2)}{2 \sqrt2 v}\Big]\rho_R\rho_L^2
\nn\\
&&-\Big[\frac{(\sqrt2 x c_{\alpha}- 2 s_{\alpha}) s_{2 \alpha}
(m_{\rho_U}^2 + 2 m_{\rho_R}^2 )}{8 v}\Big]
\rho_U\rho_R^2-\Big[\frac{x s_{\alpha}(m_{\rho_U}^2 + 2
m_{\rho_L}^2)}{2 \sqrt2 v}\Big]\rho_U \rho_L^2. \label{tril} \eea
There are no $\rho_L^3$, $\rho_U^2\rho_L$, $\rho_R^2\rho_L$, and
$\rho_U\rho_R\rho_L$ terms. The coefficient of the $\rho_U^3$
term in the $x\to 0$ limit, taking $f,\lambda,f_3,h$ finite,
reproduces the result given in \cite{Casalbuoni:2002fh}.

Concerning the gauge sector,
in the limit of large new vector boson masses,  one gets back  the
SM with the following redefinition of the gauge coupling
constants $ g^{-2}= g_0^{-2}+g_2^{-2}$,
$g'^{-2}=g_1^{-2}+g_2^{-2}$, while for the electric charge the standard
relation  $e^{-2}=g^{-2}+g'^{-2}$ holds. The fields ${V}^\pm_R$ turn
out to be unmixed and their mass is given by 
$$ M_{V_R^{\pm}}^2=
\frac{g^2 v^2}{4 s_\varphi^2 x^2}  \equiv M^2 $$ 
with
$\varphi$ defined by the relation  $g = g_2 s_{\varphi}$.
 The parameter $M$ represents the scale of  the $V_{L,R}^{\pm}~$,
$V^{}_{3L,3R}$ gauge boson masses. The standard gauge boson masses
receive corrections, due to mixing, which for $M \gg M_Z$ are of
the order $O(x^2s_\varphi^2)$. The photon is exactly massless.

The fermionic couplings of a generic gauge boson $V$ are given in
\cite{Casalbuoni:1996wa}. The heavy gauge bosons are coupled to
fermions only through mixing with the SM ones: 
for $s_\varphi \to 0$ these couplings vanish. In the following we
will use the notation $g_{V^0f\bar f}^{V(A)}$ to indicate the
vector (axial-vector) couplings of the  $V^0 = Z, V_{3L}, V_{3R}$
gauge bosons to fermions  and $g_{V^\pm f_1f_2}$ for
 charged $V^\pm = W^\pm, V_{L}^\pm, V_{R}^\pm$.

By expressing the gauge and Higgs fields  in terms of the
corresponding mass eigenstates, we derive the Higgs-gauge sector
interactions  (in the $s_\varphi x \ll 1$ limit). For the calculations
involved in this paper we will need the  following trilinear
interaction terms: 
\bea {\cal
L}^{tril}_{\rm Higgs-gauge}&\sim & \frac{v}{2} g^2  c_\alpha (1 -
2 s_\varphi^4 x^2 ) \rho_U W^{{+}} W^{{-}} +\frac{v}{2} g^2
s_\alpha (1 - 2 s_\varphi^4x^2 )\rho_R W^{{+}} W^{{-}}
\nonumber \\
&+& \frac{v}{4 c_\theta^2} g^2 c_\alpha [ 1 - 2 x^2
\frac{s_\varphi^4}{c_\theta^4} (1 - 2 c^2_\theta s^2_\theta)]
\rho_U Z^{} Z^{}\nn\\
    &+& \frac{v}{4 c_\theta^2} g^2 s_\alpha [ 1 - 2 x^2
\frac{s_\varphi^4}{ c_\theta^4} (1 - 2 c^2_\theta s^2_\theta)]
\rho_R Z^{} Z^{}\nn\\
&-& \frac{g^2 v s_\varphi}{4 c_\varphi} \{\sqrt2 x s_\alpha + 2 c_\alpha
[1 +  x^2 s_\varphi^2 (1 - 2 s_\varphi^2) ]\}\rho_U  W^{{+}} V_{L}^{{-}} \nonumber \\
&+& \frac{g^2 v s_\varphi}{4 c_\varphi} \{\sqrt2 x c_\alpha - 2
s_\alpha[1 +  x^2 s_\varphi^2 (1 - 2 s_\varphi^2) ]\}\rho_R
W^{{+}} V_{L}^{{-}}\nn\\
&-& \frac{g^2 v c_\alpha s_\varphi}{2 c_\theta c_\varphi} \left[
1 + \frac{s_\alpha x} {\sqrt{2} c_\alpha } - \right.
\left. s_\varphi^2 x^2 \frac{(2 c_\theta^2 - 1)(1 - 2
c_\theta^2 s_\theta^2)s_\varphi^2 -
c_\theta^6 c_\varphi^2}{c_\theta^4(2 c_\theta^2 - 1)}\right ]\rho_U  Z V_{3L} \nn\\
&+&\frac{g^2 v c_\alpha s_\varphi s_\theta^2 }{2 \sqrt{P} c_\theta^2}
\left[ 1 + \frac{s_\alpha x }{\sqrt{2} c_\alpha } - \right.
\left. s_\varphi^2 x^2 \frac{(2 c_\theta^2 - 1)(1 - 2 c_\theta^2 s_\theta^2)s_\varphi^2 +
s_\theta^4 P}{c_\theta^4(2 c_\theta^2 - 1)}\right ]\rho_U  Z V_{3R} \nn\\
&-&  \frac{g^2 v s_\alpha s_\varphi}{2 c_\theta c_\varphi} \left[ 1 - \frac{c_\alpha x}
{\sqrt{2} s_\alpha } - \right.
\left. s_\varphi^2 x^2 \frac{(2 c_\theta^2 - 1)(1 - 2 c_\theta^2 s_\theta^2)s_\varphi^2 -
c_\theta^6 c_\varphi^2}{c_\theta^4(2 c_\theta^2 - 1)}\right ] \rho_R  Z V_{3L}\nn\\
&+&  \frac{g^2 v s_\alpha s_\varphi s_\theta^2 }{2 \sqrt{P} c_\theta^2}
\left[ 1 - \frac{c_\alpha x }{\sqrt{2} s_\alpha } - \right.
\left. s_\varphi^2 x^2 \frac{(2 c_\theta^2 - 1)(1 - 2 c_\theta^2 s_\theta^2)
s_\varphi^2 + s_\theta^4 P}{c_\theta^4(2 c_\theta^2 - 1)}\right ]\rho_R  Z V_{3R}\nn  \\
&+&   \frac{g^2 v x s_\varphi}{2 \sqrt{2} c_\theta
c_\varphi}\rho_L Z V_{3L} +  \frac{g^2 v x s_\theta^2
s_\varphi}{2 \sqrt{2} \sqrt{P} c_\theta^2}\rho_L  Z V_{3R}\nn\\
&-& \frac{g^2 v}{4 c_\varphi^2 s_\varphi^2 x}[\sqrt{2} s_\alpha - 2 x c_\alpha s_\varphi^4 (1 + 2 c_\varphi^2 s_\varphi^2 x^2)]\rho_U V_L^+ V_L^- \nn\\
&+& \frac{g^2 v}{4 c_\varphi^2 s_\varphi^2 x}[\sqrt{2} c_\alpha + 2 x s_\alpha s_\varphi^4 (1 + 2 c_\varphi^2 s_\varphi^2 x^2)]\rho_R V_L^+ V_L^-
\label{coupgauge}
\end{eqnarray}
where $\tan\theta=g'/g$, $ P = c_{\theta}^2 - s_{\varphi}^2
s_{\theta}^2$, and we have taken only terms up to $O(
 s_\varphi^2 x^2)$ order. The couplings of $\rho_L$ to the 
light gauge bosons
are of order $ O(s_\varphi^3 x^3)$.

It is interesting to notice that the following sum rules hold:
\begin{eqnarray}
\label{sumrules}
g^2_{\rho_{U} W W} + g^2_{\rho_{R} W W} &\sim&  \frac{v^2 g^4}{4}(1 - 4 s_{\varphi}^4 x^2)\nn  \\
g^2_{\rho_{U} Z Z} + g^2_{\rho_{R} Z Z} &\sim&  \frac{v^2 g^4}{4
c_\theta^4}  [ 1 - 4 x^2  \frac{s_\varphi^4}{ c_\theta^4} (1 - 2
c^2_\theta s^2_\theta)]
\end{eqnarray}
where, for example,  we have indicated with $g_{\rho_U W W}$ the
coupling for the  $\rho_U W^+ W^-$ vertex.

\section{Scalar sector: widths and cross sections}
\label{section3} 
 Let us   evaluate the decay
partial widths and the production cross sections for the scalar
bosons $\rho_U$ and $\rho_R$ at  future LC's. Some
of the decay widths can be simply obtained by rescaling the SM
Higgs couplings   by suitable factors.  For the
$\rho_U$ boson, from  the couplings in eq. (\ref{coupgauge}) and the 
fermion couplings, we have
the following tree level partial widths, which are the relevant
ones
for the following discussions
{\begin{eqnarray}\label{larghe:rhoRU:ptreali:in}
{\Gamma (\rho_U \rightarrow \bar{f} f)} &{=}& c_\alpha^2 {\frac{N_c m_f^2}{8 \pi v^2}  m_{\rho_U} (1 - \frac{4 m_f^2}{m_{\rho_U}^2})^{3/2}}, \\
{\Gamma (\rho_U \rightarrow W W)} &\sim & c_\alpha^2 {\frac{g^4 v^2 m_{\rho_U}^3 }{256 \pi M_W^4}}  (1 - 4 x^2 s_\varphi^4)  {G(\frac{4 M_W^2}{m_{\rho_U}^2})}, \nonumber \\
{ \Gamma (\rho_U \rightarrow Z Z)} &\sim& c_\alpha^2 {\frac{g^4 v^2  m_{\rho_U}^3 }{ 512 \pi c_\theta^{4} M_Z^4}} [1 - 4 x^2  \frac{s_\varphi^4 }{c_\theta^4} (1 - 2 c_\theta^2 s_\theta^2)] { G(\frac{4 M_Z^2}{m_{\rho_U}^2})}, \nonumber
\end{eqnarray}}
where $G(z)  =  \sqrt{1 - z}(1 - z + 3 z^2/4)$ and $N_c$ is 1 for
leptons and 3 for quarks. In the subsequent numerical analysis, when computing the Higgs decay in quarks, we have used the leading
log running mass \cite{Kane:Hunter:Guide}.

When the propagator of a new vector field
 predicted by the L-BESS  model is involved in
the reactions, the rescaling of the SM formulas is no more
possible and an explicit calculation of the decay widths  and  the
cross sections becomes necessary. A first example is given by the
Higgs decay in $Wf_1 f_2$, which is relevant  for $M_W \leq
m_{\rho_{U}} \leq 2M_W$, and which has an additional Feynman
diagram
 with the contribution of the virtual $V_L$.
The final result of this  computation is
{\begin{eqnarray}\label{largh:h2wff:3}
 \Gamma_{\rho_U \rightarrow W f_1 f_2} &=&  \int_{2z}^{1+z^2}d\epsilon  \frac{ m_{\rho_U}  }
 {1536 \pi^3 M_{W_{}}^2} \Bigg( \frac{g_{\rho_{U}WW}g_{Wf_1f_2}}{1 - \epsilon} +
 {\frac{g_{\rho_{U}WV_L}g_{V_Lf_1f_2}}{1 - \epsilon + z^2 - \frac{M_{V_L}^2}{m_{\rho_U}^2}}}
 \Bigg)^2  \times \nonumber \\
&\times & \sqrt{\epsilon^2 - 4 z^2 }(8 z^2 - 12 z^2 \epsilon + 12 z^4 + \epsilon^2)
\end{eqnarray}}
where $\epsilon = 2 E_W/m_{\rho_U}$, $ z = M_{W}/m_{\rho_U}$. The
couplings $g_{\rho_{U}W W}$, $g_{\rho_{U}WV_L}$ can be extracted  
from eq. (\ref{coupgauge})
and the fermionic couplings  $g_{W f_1 f_2}$
and $g_{V_L f_1 f_2}$ from \cite{Casalbuoni:1996wa}.

Concerning the  partial widths for the heavy Higgs $\rho_R$, these  are obtained by simply replacing $c_\alpha$ with $s_\alpha$ in eq. (\ref{larghe:rhoRU:ptreali:in}).
However if
 $m_{\rho_R} \geq 2m_{\rho_U}$ the new  decay $\rho_R \rightarrow \rho_U \rho_U$
is allowed; the corresponding width, using the trilinear coupling
 in eq. (\ref{tril}),  is given by 
{\be \Gamma ( \rho_R
\rightarrow \rho_U \rho_U) = \frac{1}{512 v^2 \pi m_{\rho_R}} (1
- \frac{4 m_{\rho_U}^2}{m_{\rho_R}^2})^{1/2} {(2
c_{\alpha} + \sqrt2 x s_{\alpha})^2 s_{2 \alpha}^2 (2 m_{\rho_U}^2 +
m_{\rho_R}^2 )^2} . 
\ee}

The main channels for $\rho_U$ or $\rho_R$ Higgs production at $e^+ e^-$ colliders are the Higgstrahlung and the fusion channel (like in the SM). In the L-BESS model these processes get additional contributions by the exchange of the new vector resonances. In the case of $\rho_U$ the Higgstrahlung $e^+ e^- \rightarrow Z^*, V_{3L}^*, V_{3R}^* \rightarrow Z \rho_U$ cross section is given by
{\begin{equation}\label{sezdurto:ee2uZ:Gtot}
 \sigma(e^+ e^- \rightarrow \rho_U Z)= 
\frac{\sqrt{\lambda_U} ({\lambda_U} +  12 m_Z^2 s)}{192 \pi m_Z^2 s^2} [(G^{V}_{TOT} )^2 + (G^{A}_{TOT} )^2]
\end{equation}}
where {$\lambda_U = (s - m_{\rho_{U}}^2 - M_Z^2)^2 - 4 m_{\rho_{U}}^2 M_Z^2$}, $\sqrt{s}$ is the center of mass energy and
{\begin{equation}
G_{TOT}^{V(A)} = \frac{g_{Zf\bar{f}}^{V(A)}g_{\rho_U Z Z}}{(s - M_Z^2)} {+ \frac{g_{V_{3L}f\bar{f}}^{V(A)}g_{\rho_U Z V_{3L}}}{(s - M_{V_{3L}}^2)} + \frac{g_{V_{3R}f\bar{f}}^{V(A)}g_{\rho_U Z V_{3R}}}{(s - M_{V_{3R}}^2)}}.
\label{couplings}
\end{equation}}
The  couplings which appear in eq. (\ref{couplings})  are
extracted  from eq. (\ref{coupgauge}) and the fermionic couplings
from \cite{Casalbuoni:1996wa}.

The $e^+ e^- \rightarrow V^{*\pm} V^{*\mp} \nu_e \bar{\nu}_e
\rightarrow  \nu_e \bar{\nu}_e \rho_{U,R}$ Higgs boson production
cross section via $V^\pm V^\mp$ fusion (with $V^\pm = W^\pm,
V_L^\pm $) has been  obtained by implementing this model in 
 the  program
COMPHEP \cite{Pukhov:1999gg}.

\section{Numerical analysis}
\label{section4} Before studying the phenomenology of the  scalar sector
of the model
at future LC's  we must fix the physical parameters of the L-BESS model. For
the new parameters $M$ and $s_\varphi$ we choose values inside
 the region allowed by the present electroweak precision data.
This region is obtained by comparing the prediction of the L-BESS model
 for the $\epsilon$ parameters 
 with their  experimental values \cite{Casalbuoni:1996wa}.
This leads, assuming the SM radiative corrections to the $\epsilon$, 
to  a $95\%$ CL bound which, for $M\gtrsim500~\gev$ reads
$M(\gev) \gtrsim 2000 ~ s_\varphi$ and slightly depends on the choice of
the SM Higgs mass. 
The contributions from the additional Higgses should be also included;
we however expect that this inclusion will not dramatically change the results
due to the sum rules in eq. (\ref{sumrules}).
We will consider the following choices
  $(M, s_\varphi) = (500\,\,\gev, 0.25)$,
 $(1000\,\,\gev, 0.5)$, $(1500\,\,\gev, 0.75)$
and as a reference, we also consider the case corresponding to
  the decoupling limit $M \rightarrow \infty$.
For the  scalar sector parameters we
  take $\alpha \in [0, \pi/2] $ and $m_{\rho_R} > m_{\rho_U}$ ($f>0$).

The $\rho_U$ and $\rho_R$ partial decay widths in $WW$, $ZZ$
  show a negligible dependence on the parameters $M$ and $s_\varphi$ chosen in the
  allowed region;
 moreover from the numerical integration of eq. (\ref{largh:h2wff:3}), we can 
show that  the contribution of
 heavy virtual
vector bosons is not appreciable.
 Therefore such decay widths are modified, with good approximation,
 by a factor $c_\alpha^2$ and $s_\alpha^2$ with respect to those of the SM  for
 $\rho_U$ and $\rho_R$ respectively.
As a consequence   the corresponding Branching Ratios (BR's)
 for $\rho_U$ are substantially similar to those of the SM (this does not happen for  the  supersymmetric Higgses away from the decoupling limit).

\begin{figure}[htbp]
\begin{center}
\includegraphics[width=10cm]{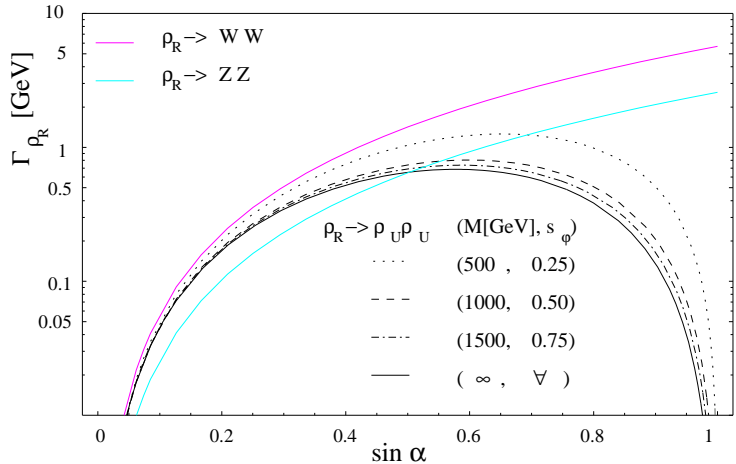}
\end{center}
\caption{\linespread{0.9}\small{The most important partial widths
for the $\rho_R$ Higgs boson decay as a function of $s_\alpha$ for
$m_{\rho_R} = 300 \,\,\gev$ and $m_{\rho_U} = 120 \,\,\gev$. The
continuous  magenta (dark gray) line corresponds to the process
$\rho_R \rightarrow W W$ and the continuous cyan (light gray)
line to $\rho_R \rightarrow Z Z$. The black lines corresponds to
the decay $\rho_R \rightarrow \rho_U \rho_U$ for $(M,s_\varphi) =
(500 \,\gev,0.25)$ (dotted line), $(M,s_\varphi) = (1000 \,\gev,
0.5)$ (dashed line), $(M,s_\varphi) = (1500 \,\gev, 0.75)$
(dash-dotted line) and $M = \infty$ (continuous line).
\label{grafLargRAlphaMvGener.eps}}}
\end{figure}

The only partial width which depends considerably on the parameters $M$
and $s_\varphi$ is the one  relative  to the decay $\rho_R
\rightarrow \rho_U \rho_U$, as shown in  Fig.
\ref{grafLargRAlphaMvGener.eps}. When kinematically allowed,
$\Gamma(\rho_R \rightarrow \rho_U \rho_U)$ is, 
for small $s_\alpha$, of the same order of the
dominant  $\Gamma(\rho_R \rightarrow W^+W^-)$.
Therefore the BR's of the $\rho_R$ boson can be different from
those of the SM  Higgs boson. Because of this possible  decay
also the total width depends on $M$ and $s_\varphi$.

\begin{figure}[h!]
\begin{center}
\begin{tabular}{c c}
\epsfig{file=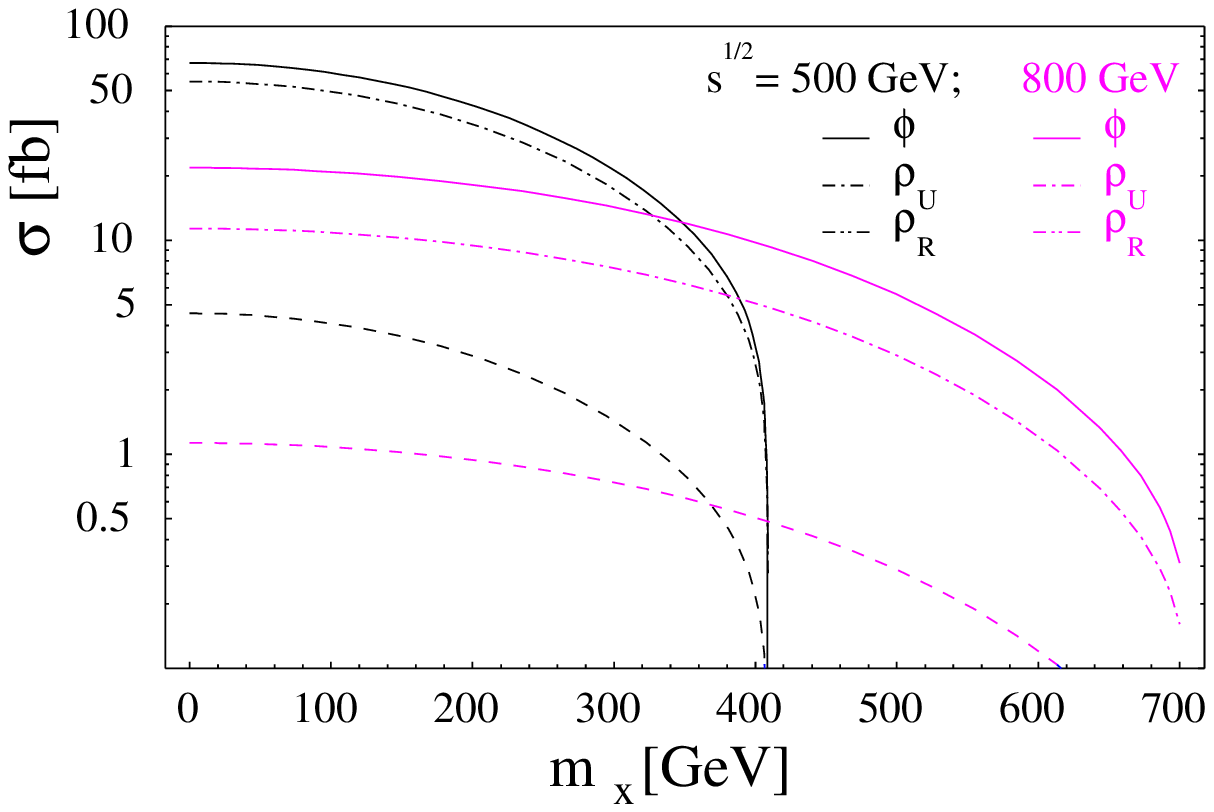
,width=7.5cm,height=6.0cm
} &
\epsfig{file=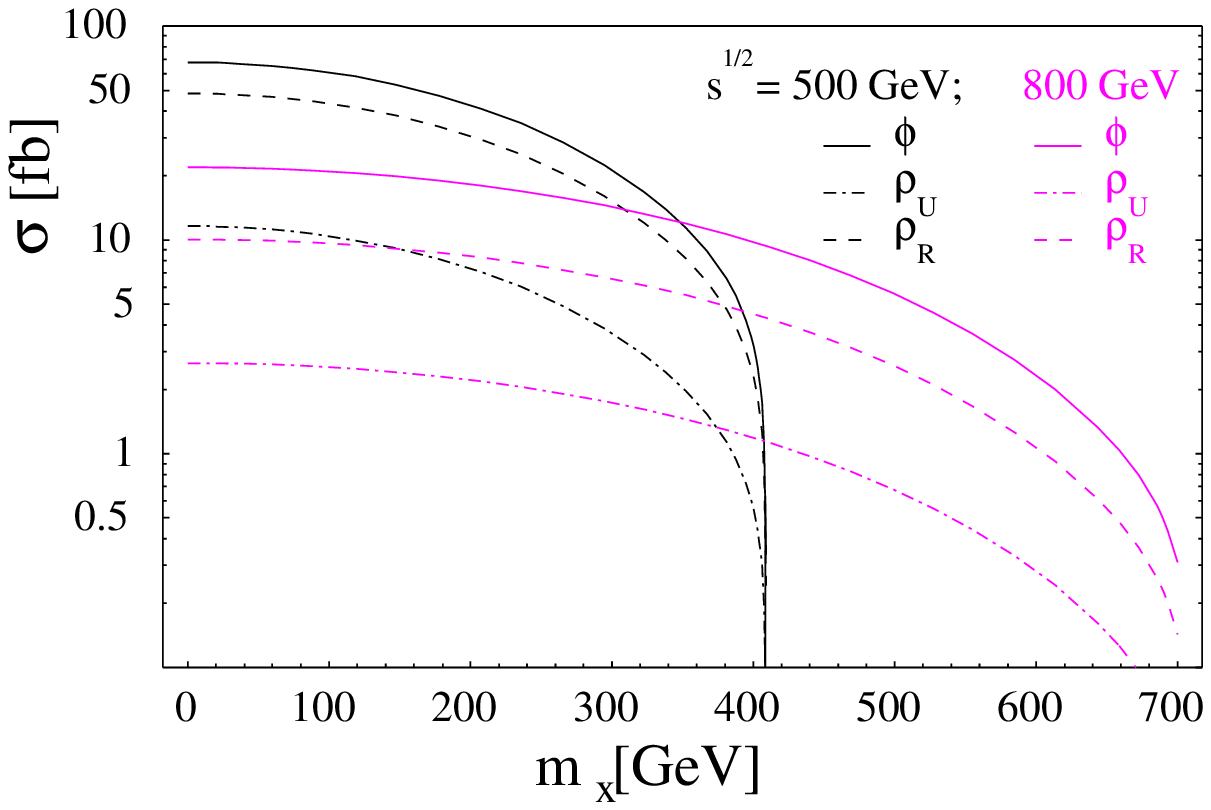
,width=7.5cm,height=6.0cm
} \\
\end{tabular}
\end{center}
\vspace*{-0.75cm} \caption{\linespread{0.9}\small{Higgstrahlung
cross sections as a  function of the Higgs boson mass $m_X$  for
the L-BESS model $\rho_U$ boson (dash-dotted lines) and $\rho_R$ boson
(dashed lines) , and for the SM $\phi$ boson (continuous lines),
with: $\sqrt{s} = 500 \, \, \gev$ (black lines) and $\sqrt{s} =
800 \, \, \gev$ (magenta (gray) lines),  $M = 1000 \, \, \gev\, $,
$s_\varphi = 0.5$, $s_\alpha = 0.25$ (left panel), $s_\alpha = 0.9$
(right panel).\label{ee2V2ZX2}}}
\end{figure}

Let us now study the production cross sections. The Higgstrahlung cross sections for the $\rho_U$ and
the $\rho_R$ are shown in Fig. \ref{ee2V2ZX2} for  $s_\alpha = 0.25$
 (left panel)
and for $s_\alpha = 0.9$ (right panel), with $M = 1000 \, \,
\gev\, $, $s_\varphi = 0.5\, \, $ at a LC with  $\sqrt{s} = 500
\, \, \gev$ (black lines), $\sqrt{s} = 800 \, \, \gev$ (magenta
(gray) lines).

In general for  $s_\alpha \gtrsim 0.7$ the production rate
for
 the $\rho_R$ can be greater than  the  $\rho_U$ one
even if $m_{\rho_R}>m_{\rho_U}$; this means that, in this case,
the heavier Higgs boson $\rho_R$
could be   detected at a LC before the lighter $\rho_U$, (see
 Fig. \ref{ee2V2ZX2}).

The Higgstrahlung cross section is sensitive to the  new vector
resonances: in fact for  center of mass energies close to $M$ the
 $\rho_U$ production via  this mechanism differs
from the SM one by a factor that can  be  much more
different from the naive $c_\alpha^2$ due to the coupling.


For increasing values of the energy  of the collider
 the $V^{\pm} V^{\mp}$ fusion process
 becomes dominant with respect to the Higgstrahlung process.
We have computed the fusion cross section by using
the code \textsc{COMPHEP} \cite{Pukhov:1999gg}: the results for the production
of the Higgs  $\rho_R$ for $\sqrt{s} = 800 \,\gev$, $s_\alpha = 0.25$,
 $(M, s_\varphi) = (1000\,\gev, 0.25)$, are shown in Table \ref{tab:ww:fus:R}
 for different values of  $m_{\rho_R}$.
For comparison we  also give the Higgstrahlung
cross section values.
For example, for an integrated luminosity of
1000 $\fbi$ and $m_{\rho_{R}}=600\gev$, one has 150 fusion events and detection
is possible, while for $m_{\rho_{R}}=700\gev$ one has only 16 events and an accurate analysis of signal to background ratio is required.

In Table \ref{tab:VV:fus:R:Alpha} we show the  fusion cross
section for the process $V^\pm V^\mp \rightarrow 
\rho_{U}$ for \break $\sqrt{s}\ =\ 500\,\gev$, $m_{\rho_U} = 120 \,\gev$, $s_\varphi =  0.5$
and
  different values of  $s_\alpha$ and
$M$. A decrease in the SM Higgs
fusion cross section can be the consequence of the
  presence of new vectors and/or  the $c_\alpha^2$ factor as in the case of
the Higgstrahlung process.

\begin{table}
\begin{center}
\begin{tabular}{|c|cccccc|}
\hline $m_{\rho_R}[\gev]$ & 100  & 200 & 300 & 400 & 500 & 600 \\
\hline \hline 
 $\sigma_{fusion V^\pm V^\mp} [\fb]$
& 12.0 & 6.73 & 3.48 & 1.60 & 0.60 & 0.15  \\
 $\sigma_{Higgstrahlung} [\fb]$& 1.32 & 1.15 & 0.90 & 0.62 & 0.35&
0.15 \\
\hline
\end{tabular}
\end{center}
\caption{\linespread{0.9}\small{
Fusion  cross section $V^\pm V^\mp$ with
$V^\pm = W^\pm, V_L^\pm$  and Higgstrahlung for different values of
$m_{\rho_R}$
with $\sqrt{s}= 800 \gev$, $s_\alpha = 0.25$
and $(M, s_\varphi)= (1000 \gev , 0.25)$.}}\label{tab:ww:fus:R}
\end{table}

\begin{table}
\begin{center}
\begin{tabular}{|c||ccc|}
\hline $ \sigma_{fusion V^\pm V^\mp }$ & $M = 1000 \gev$ & $M = 2000
\gev$ & $M = \infty$ \\
\hline \hline $s_\alpha = 0$ & $75.6 \fb$ & $77.4 \,\,fb$ &
$77.9\,\,fb$ \\
$s_\alpha = 0.25$ & $70.7 \fb$ & $72.3\,\, fb$ & $73.1 \,\,fb$ \\
$s_\alpha = 0.5$ & $56.2\fb$  & $57.9\,\, fb$ & $58.4\,\,fb$ \\
\hline
\end{tabular}
\end{center}
\caption{\linespread{0.9}\small{Fusion cross section $V^\pm V^\mp\to
 \rho_{U}$ with $M = 1000 \,\, GeV$, $M = 2000 \gev$
and $M = \infty$ for $s_\alpha = 0$, $s_\alpha = 0.25$,  $s_\alpha =
0.5$;
$\sqrt{s}= 500 \gev$, $s_\varphi = 0.5$, $m_{\rho_{U}}=120
\gev$.}}\label{tab:VV:fus:R:Alpha}
\end{table}

\begin{figure}[htbp]
\begin{center}
\includegraphics[width=7cm]{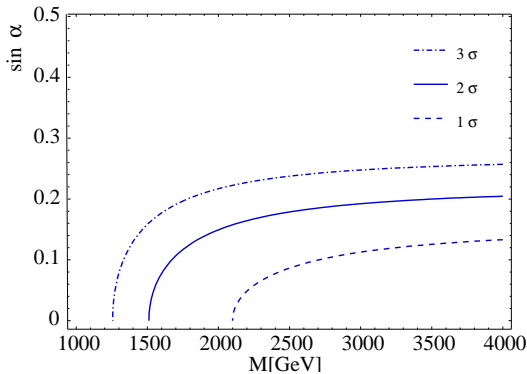}
\end{center}
\caption{\linespread{0.9}\small{$1\sigma$ (dashed line), $2\sigma$
(continuous line) and  $3\sigma$ (dash-dotted line) contours in
the plane ($M,s_\alpha$) from deviations in the   $\rho_U$
Higgstrahlung  with respect to the SM for   $ m_{\rho_U} = 120
\gev$, $s_\varphi = 0.5$, $\sqrt{s} = 500\gev$ and $L=500$ fb$^{-1}$.}}
\label{cntrDeltaSigmaUZ.eps}
\end{figure}

\begin{table}
\begin{center}
\begin{tabular}{|c|cccccc|}
\hline  & $g^2_{\rho_U WW}$ & $g^2_{\rho_U Z Z}$ & $g^2_{\rho_U b b}$
& $g^2_{\rho_U \tau \tau}$ & $g^2_{\rho_U c c}$ & $g^2_{\rho_U t t}$  \\
\hline  $(\Delta g^2/g^2)_{ex}$ & 2.4\% & 2.4\% & 4.4\% & 6.6\% & 7.4\% &  10\% \\
\hline  $(\Delta g^2/g^2)_{th}$ & - & -  & 3.5\% & - & 24\% &  2.5\% \\
\hline  $s_\alpha$ & 0.22 & 0.22 & 0.34 & 0.36 & 0.71 &  0.45 \\
\hline
\end{tabular}
\end{center}
\caption{\linespread{0.9}\small{$s_\alpha$ upper bounds assuming a $2 \sigma$
deviation (with respect to the SM prediction)
in the measurements of squared couplings  of a lightest Higgs
$\rho_U$, with $ m_{\rho_U}=120 \gev $, to $W W^*$, $ZZ^*$, $b \bar{b}$, $\tau^+ \tau^- $, $c \bar{c}$ and $t \bar{t}$ assuming
  $\sqrt{s} = 500\gev$, $M >> 4000 \gev$, $s_\varphi << 0.5$ and an
integrated luminosity of $500 \fbi$ except for $tt$ ($\sqrt{s} = 800\gev$ and
$L=1000\fbi$ from $e^+e^-\to \bar ttH$.)}}
\label{tab:riass:soglie:largh}
\end{table}

\section{Bounds from precision measurements at future LC}
\label{section6}

A LC besides detecting  one or more Higgs bosons can
also determine with precision their masses, their couplings to fermions
and to gauge bosons and their trilinear couplings. In the scenario where
only one light Higgs boson has been discovered these precision measurements
allow to get bounds on extended electroweak models like the one we are
considering.
We have assumed an experimental uncertainty on the determination
of the Higgstrahlung cross section $\Delta \sigma/\sigma=2.4\%$ \cite{Battaglia:2000jb}.
If  no deviation  with respect to the prediction of the SM is 
observed on  $\sigma_{Higgstrahlung}$, at a LC with $\sqrt{s} = 500\gev$ and $L=500$ 
fb$^{-1}  $,
 one  gets the bounds in the plane ($M,s_\alpha$),  shown in Fig. \ref{cntrDeltaSigmaUZ.eps}.
For example, for $M \sim 4\tev$, so that the new resonances are not accessible at the
LHC, the 95\% C.L.limit  on $s_\alpha$ is $ \sim 0.2$.


By combining the fusion cross section, the Higgstrahlung cross section,
the measurements of different Higgs branching fractions and
$e^+e^-\to \bar ttH$ cross section
one can extract
the Higgs squared couplings
to fermions and gauge bosons or equivalently the partial widths with
the experimental uncertainty given in \cite{Battaglia:2000jb,Conway:2002kk}.
Assuming
no deviations with respect to the SM, the  $2\sigma$ upper bounds
on $s_\alpha$ are given in Table \ref{tab:riass:soglie:largh}.
In deriving these limits we have also taken into account the theoretical uncertainties,
also given in
 Table \ref{tab:riass:soglie:largh}.
The strongest bounds come from the measurements of
$g^2_{\rho_U WW}$,  $g^2_{\rho_U Z Z}$ and $g^2_{\rho_U b b}$.

LC measurements of double Higgs production
$e^+e^-\to HHZ$ and $e^+e^-\to\nu_e\bar\nu_e HH$
can also determine the Higgs trilinear coupling for
Higgs masses in the range $120-180~GeV$ with an accuracy of
 22\%  (for $\sqrt{s}=500\gev$ and $L=1000\fbi$) \cite{Battaglia:2000jb}
 and  of  8\% (for a multiTev LC) \cite{Battaglia:2001nn}. Using the
expression given in eq. (\ref{tril}) this last bound can be
translated in a 2$ \sigma$ limit on $s_\alpha\sim 0.33$.

\section{Conclusions}
We have discussed the scalar sector of the linearized version of the BESS model which predicts three scalar states:
 $\rho_U$, $\rho_R$ and $\rho_L$. The $\rho_U$ and $\rho_R$ bosons mix 
 and therefore, depending on the mixing angle, can
be detected  at the LHC and at a LC, instead the $\rho_L$ has no
coupling to fermions and suppressed couplings to SM gauge bosons.
At the LHC the best channel for an heavy Higgs is the $ZZ\to  4\ell$
channel, while at a LC the recoil technique allows the discovery
no matter how the $\rho_R$ decays. The main
decay channels of the $\rho_R$ are $WW$ and $\rho_U \rho_U$
(when kinematically allowed and for small $s_\alpha$): 
therefore detection of $\rho_R$ at a LC is
possible in a larger region of the parameter space. The LC's offer,
in addition to the detection of the scalar particles,  the
possibility of discriminating among different models by accurate
measurements of the production cross sections and the Higgs
couplings, by combining measurements of branching ratios, 
Higgstrahlung and fusion cross sections.

\begin{thebibliography}{99}
\footnotesize

\bibitem{Schmaltz:2002wx}
M. Schmaltz.
\newblock 
Physics beyond the standard model (theory): Introducing the little  Higgs.
\newblock {\em Nucl. Phys. Proc. Suppl.}, 117:40--49, 2003.

\bibitem{Chivukula:1996gu}
R.~S. Chivukula, E.~H. Simmons, and J.~Terning.
\newblock {Limits on noncommuting extended technicolor}.
\newblock {\em Phys. Rev.}, D53:5258--5267, 1996;
R.~S. Chivukula, E.~H. Simmons, J. Howard, and H.-J. He.
\newblock Precision electroweak constraints on hidden local symmetries.
\newblock hep-ph/0304060.

\bibitem{Casalbuoni:1995yb}
R.~Casalbuoni et~al.
\newblock Symmetries for vector and axial vector mesons.
\newblock {\em Phys. Lett.}, B349:533--540, 1995;
R.~Casalbuoni et~al.
\newblock 
Low energy strong electroweak sector with decoupling.
\newblock {\em Phys. Rev.}, D53:5201--5221, 1996.

\bibitem{Casalbuoni:1996wa}
R.~Casalbuoni, S.~De~Curtis, D.~Dominici, and M.~Grazzini.
\newblock {An extension of the electroweak model with decoupling at low
  energy}.
\newblock {\em Phys. Lett.}, B388:112--120, 1996;
R.~Casalbuoni, S.~De~Curtis, D.~Dominici, and M.~Grazzini.
\newblock {New vector bosons in the electroweak sector: A renormalizable model
  with decoupling}.
\newblock {\em Phys. Rev.}, D56:5731--5747, 1997.

\bibitem{Casalbuoni:2000gn}
R.~Casalbuoni, S.~De~Curtis, and M.~Redi.
\newblock Signals of the degenerate {BESS} model at the {LHC}.
\newblock {\em Eur. Phys. J.}, C18:65--71, 2000.

\bibitem{Casalbuoni:2002fh}
R.~Casalbuoni and L.~Marconi.
\newblock The linear BESS model and the double Higgs-strahlung production.
\newblock {\em J. Phys.}, G29:1053--1060, 2003.



\bibitem{Kane:Hunter:Guide}
{G. F. Gunion, E. H. Howard,  G. Kane and S. Dawson}.
\newblock {\em {The Higgs Hunter's Guide}}.
\newblock {Perseus Publishing}, 2000.

\bibitem{Pukhov:1999gg}
A.~Pukhov et~al.
\newblock {COMPHEP: A package for evaluation of Feynman diagrams and
  integration over multi-particle phase space. User's manual for version 33}.
\newblock 1999.

\bibitem{Battaglia:2000jb}
M.~Battaglia and K.~Desch.
\newblock {Precision studies of the Higgs boson profile at the $e^+ e^-$ Linear
  Collider}.
\newblock hep-ph/0101165.
\newblock Batavia 2000, Physics and experiments with future linear $e^+ e^-$
 colliders 163-182, 2000.

\bibitem{Conway:2002kk}
J.~Conway, K.~Desch, J.~F. Gunion, S.~Mrenna, and D.~Zeppenfeld.
\newblock The precision of Higgs boson measurements and their implications.
\newblock {\em eConf}, C010630:P1WG2, 2001.

\bibitem{Battaglia:2001nn}
M. Battaglia, E. Boos, and W.-M. Yao.
\newblock Studying the Higgs potential at the $e^+ e^-$ Linear Collider.
\newblock {\em eConf}, C010630:E3016, 2001.

\end{thebibliography}

\vspace*{0.75cm} \noindent We wish to thank M. Battaglia for
interesting discussions.

\end{document}